\documentclass{Interspeech2024}
\usepackage{url}
\usepackage{booktabs,subfigure,float,amsmath,amssymb}
\usepackage{adjustbox}
\usepackage{threeparttable}
\usepackage[para]{footmisc}
\usepackage{lipsum}
\usepackage{tablefootnote}
\usepackage{multirow}
\usepackage{ulem}
\usepackage{xcolor}

\interspeechcameraready

\title{GLOBE: A High-quality English Corpus with Global Accents for Zero-shot Speaker Adaptive Text-to-Speech}
\name{Wenbin Wang$^1$, Yang Song$^1$, Sanjay Jha$^1$}

\address{
  $^1$School of Computer Science and Engineering, University of New South Wales, Australia}
\email{wenbin.wang@unsw.edu.au, yang.song1@unsw.edu.au, sanjay.jha@unsw.edu.au}


\newcommand{\red}[1]{\textcolor{black}{#1}}
\newcommand{\blue}[1]{\textcolor{black}{#1}}
\newcommand{\rec}[1]{\textcolor{black}{#1}}
\newcommand{\reg}[1]{\textcolor{black}{#1}}

\begin{document}

\maketitle

\begin{abstract}
    
This paper introduces GLOBE, a high-quality English corpus with worldwide accents, specifically designed to address the limitations of current zero-shot speaker adaptive Text-to-Speech (TTS) systems that exhibit poor generalizability in adapting to speakers with accents. Compared to commonly used English corpora, such as LibriTTS and VCTK, GLOBE is unique in its inclusion of utterances from 23,519 speakers and covers 164 accents worldwide, along with detailed metadata for these speakers. Compared to its original corpus, i.e., Common Voice, GLOBE significantly improves the quality of the speech data through rigorous filtering and enhancement processes, while also populating all missing speaker metadata. The final curated GLOBE corpus includes 535 hours of speech data at a 24 kHz sampling rate. Our benchmark results indicate that the speaker adaptive TTS model trained on the GLOBE corpus can synthesize speech with better speaker similarity and comparable naturalness than that trained on other popular corpora. We will release GLOBE publicly after acceptance. \reg{The GLOBE dataset is available at \url{https://globecorpus.github.io/}}.
\end{abstract}

\section{Introduction}

Recent advances in deep learning have significantly propelled TTS research forward. The latest generation of neural networks-based TTS models \cite{DBLP:conf/iclr/0006H0QZZL21, DBLP:conf/icml/PopovVGSK21, DBLP:conf/icml/KimKS21, DBLP:conf/mm/YeXTCLG23} can now generate highly lifelike human speech. This advancement has shifted the TTS research focus toward more sophisticated and challenging tasks \cite{DBLP:journals/corr/abs-2106-15561}. Among these emerging tasks, speaker adaptive TTS \cite{DBLP:journals/corr/abs-2106-15561, DBLP:conf/iclr/Chen0LLQZL21}, also known as voice cloning, especially in zero-shot scenarios \cite{DBLP:conf/icml/CasanovaWSJGP22, gzs, DBLP:journals/corr/abs-2306-03509}, has emerged as an active area of interest. \textit{Zero-shot speaker adaptive TTS} allows TTS models to swiftly adapt to new speaker voices, which are not included in the training dataset, using only seconds of speech samples. This technique significantly broadens TTS technology's acceptability.

\reg{ In our previous works \cite{gzs, DBLP:journals/taslp/WangSJ24}, we found} a significant challenge in current zero-shot speaker adaptive TTS research is models' limited generalizability to accented voices. Despite increasing model parameters and enlarging the training dataset, this challenge persists \cite{DBLP:journals/corr/abs-2301-02111, DBLP:journals/corr/abs-2306-03509}. Our analysis identifies one of the crucial factors contributing to this issue: the prevalent English TTS datasets contain a limited set of accents. For example, LibriTTS \cite{DBLP:conf/interspeech/ZenDCZWJCW19} and LibriSpeech \cite{DBLP:conf/icassp/PanayotovCPK15} mainly consists of speakers with US English during the filtering process, and the VCTK \cite{VCTK} dataset encompasses speakers with only 11 accents. The Common Voice dataset \footnote{In this paper, the ``Common Voice'' dataset specifically refers to its Common Voice Corpus 14.0 English subset at https://commonvoice.mozilla.org/en/datasets. The original Common Voice dataset includes multiple languages and versions.} \cite{DBLP:conf/lrec/ArdilaBDKMHMSTW20}, \reg{which comprises more than 3,000 hours of speech and covers up to 337 accents,} presents a potential solution. However, it also exhibits several undesirable characteristics for building TTS system \cite{DBLP:conf/slt/OgunCV22}: 1) A significant number of speech samples contain noticeable background or electromagnetic noise; 2) Despite the audio file has a sample rate of 48 kHz, the actual signal bandwidth is limited; 3) \reg{Many samples feature mispronunciations or corrections where speakers repeat unfamiliar words upon realizing a mispronunciation;} 4) Half of the speakers have missing metadata and the accent labels are confusing.

To address these issues, we introduce GLOBE, a hi\textbf{G}h-quality eng\textbf{L}ish c\textbf{O}rpus with glo\textbf{B}al acc\textbf{E}nts, based on the Common Voice dataset. To construct this data, we remove low-quality, bandwidth-limited audio samples and re-align the utterance and text. Then, we manually cleaned the accent labels and populated missing speaker metadata through our prediction model. Compared to other popular English TTS datasets \cite{VCTK, DBLP:conf/interspeech/ZenDCZWJCW19, DBLP:conf/lrec/ArdilaBDKMHMSTW20}, the GLOBE dataset has its unique features:

\textbf{High Speech Quality:} The GLOBE dataset contains 535 hours of high-quality speech filtered from over 3,000 hours, with signal-to-noise ratio, signal bandwidth, and transcription accuracy. Our experimental results on zero-shot speaker adaptive TTS indicate that speech samples in GLOBE surpass those in VCTK and LibriTTS in terms of objective and subjective naturalness in mean opinion score evaluations.

\textbf{Global Accent Coverage:} With 23,519 speakers representing 164 different English accents from more than 50 countries, GLOBE offers unparalleled accent diversity. Our experiments demonstrate that such diversity significantly improves the generalizability of zero-shot speaker adaptive TTS models to different accents.

\textbf{Extra Speaker Information:} In addition to speech audio and corresponding text, GLOBE provides detailed metadata for all 23,519 speakers, including accent, age, and gender. This additional information enables future research of more personalized TTS models and mitigation of bias.

\begin{table*}[h]
\caption{Statistics on GLOBE and Relevant English Multi-speaker Corpus}
\begin{adjustbox}{width=15cm,center}
\label{summerize}
\scriptsize
\begin{tabular}{c|cccccc}
\toprule
Corpus     & Total Hours & Total Speakers   &Sample Rate     & Total Accent    & Speaker Info   & License   \\ \midrule
CSTR VCTK \cite{VCTK}  & 44    & 109          &   48 kHz    &   11        & Accent, Gender & CC BY 4.0 \cite{CC4} \\ 
LibriTTS \cite{DBLP:conf/interspeech/ZenDCZWJCW19}  & 586   & 2,456        &   24 kHz     &   -       & - & CC BY 4.0 \cite{CC4} \\ 
LibriTTS-R \cite{DBLP:journals/corr/abs-2305-18802} & 585   & 2,456         &   24 kHz    &   -       & - & CC BY 4.0 \cite{CC4} \\ 
Common Voice \cite{DBLP:conf/lrec/ArdilaBDKMHMSTW20} & 3,347   & 88,904  &   48 kHz  &   337          & Accent, Age, Gender & CC-0 1.0 \cite{CC0} \\ 
GLOBE       & 535   & 23,519        &   24 kHz    &   164       & Accent, Age, Gender & CC-0 1.0 \cite{CC0}\\ \bottomrule 
\end{tabular}
\end{adjustbox}
\end{table*}

\section{Relevant English Multi-speaker Corpus}
\label{datasets}
\quad\quad\textbf{VCTK \cite{VCTK}.} The VCTK dataset is a widely utilized corpus for developing TTS and voice cloning systems. It contains 44 hours of 48 kHz speech data from 109 English speakers and each speaker reads about 400 sentences from newspapers. Moreover, the VCTK dataset includes labels for each speaker's accent and gender, covering a total of 11 accents. 

\textbf{LibriTTS \cite{DBLP:conf/interspeech/ZenDCZWJCW19}.} The LibriTTS dataset is a well-known multi-speaker dataset for training speaker adaptive TTS systems. It derives from the LibriSpeech dataset \cite{DBLP:conf/icassp/PanayotovCPK15} and includes 585 hours of audio recordings at a 24 kHz sampling rate, contributed by 2,456 speakers. This dataset specifically targets and mitigates various limitations in the original LibriSpeech collection, making it suitable for TTS system applications.

\textbf{LibriTTS-R \cite{DBLP:journals/corr/abs-2305-18802}.} The LibriTTS-R corpus, an enhanced version of the LibriTTS dataset, significantly improves audio quality by incorporating speech restoration techniques. These enhancements support the training of high-quality TTS models. This corpus is the same as LibriTTS, retaining 585 hours of speech data from 2,456 speakers.

\textbf{Common Voice \cite{DBLP:conf/lrec/ArdilaBDKMHMSTW20}.} Common Voice is a dataset powered by the voices of volunteer contributors from all around the world. It contains 3,347 hours of audio from 88,904 speakers, recorded at a 48kHz sample rate. The Common Voice dataset is frequently utilized to build automatic speech recognition systems. Another study \cite{DBLP:conf/slt/OgunCV22} demonstrates that despite Common Voice's extensive collection of audio, it is not suitable for building TTS models due to the prevalence of poor-quality audio.

\rec{The key information for these datasets, along with GLOBE, is summarized in Table 1.}

\section{Data Processing Pipeline}

\subsection{Speech Sample Pre-processing and Filtering}

During the initial construction phase of the GLOBE corpus, low-quality speech samples are identified and removed to prevent adverse impacts on the performance of TTS models when utilized as training data. The primary metric used for assessing speech sample quality is the signal-to-noise ratio (SNR), estimated through waveform amplitude distribution analysis \cite{DBLP:conf/interspeech/KimS08}, as a crucial indicator of audio clarity by quantifying the ratio of unwanted noise to identifiable speech. In line with \cite{DBLP:conf/interspeech/ZenDCZWJCW19}, utterances with an SNR below 0 dB are excluded from the GLOBE corpus. Furthermore, the actual signal bandwidth is determined by identifying the highest frequency that is at least -50 dB below the power spectrogram’s peak average, following the methodology outlined in \cite{DBLP:conf/interspeech/BakhturinaLGZ21}. Utterances with actual signal bandwidths below 12 kHz are excluded. Additionally, utterances containing more than 930 milliseconds of continuous internal silence, attributed to abnormal pauses or hesitations, are removed to further avoid negative impacts on the duration predictors within TTS models. The internal silence is detected by a voice activity detection (VAD) tool \footnote{https://github.com/snakers4/silero-vad} and the selection of the 930-millisecond threshold is derived from the maximum duration of internal silence observed within the LibriTTS clean subset.

\subsection{Speech Text Alignment}

\blue{Given the significantly higher word error rate of the Common Voice dataset compared to other popular TTS datasets \cite{VCTK, DBLP:conf/interspeech/ZenDCZWJCW19}, which seriously impacts the intelligibility of synthesized speech if trained on these data \cite{DBLP:conf/slt/OgunCV22}, the next phase in the development of the GLOBE corpus focuses on improving transcription accuracy. Following \cite{DBLP:conf/interspeech/ZenDCZWJCW19},} Whisper \cite{DBLP:conf/icml/RadfordKXBMS23} is firstly utilized to transcribe all utterances and a weighted finite-state transducer-based system \cite{DBLP:conf/interspeech/ZhangBG21} is then employed to normalize the corresponding text from the original dataset and the transcribed texts to their spoken forms. Following that, the word-level edit distance between them is computed by a publicly available toolkit \footnote{https://github.com/roy-ht/editdistance}. Utterances exhibiting an edit distance greater than 1 are eliminated. Furthermore, clips that have an edit distance of 1 due to consecutive word repetitions are also discarded, \blue{as this often results from the repeated pronunciation of unfamiliar words according to our analysis.}

\subsection{Speaker Information Refinement and Speech Post-processing}

\begin{figure}[t]
\centering
\includegraphics[width=0.45\textwidth]{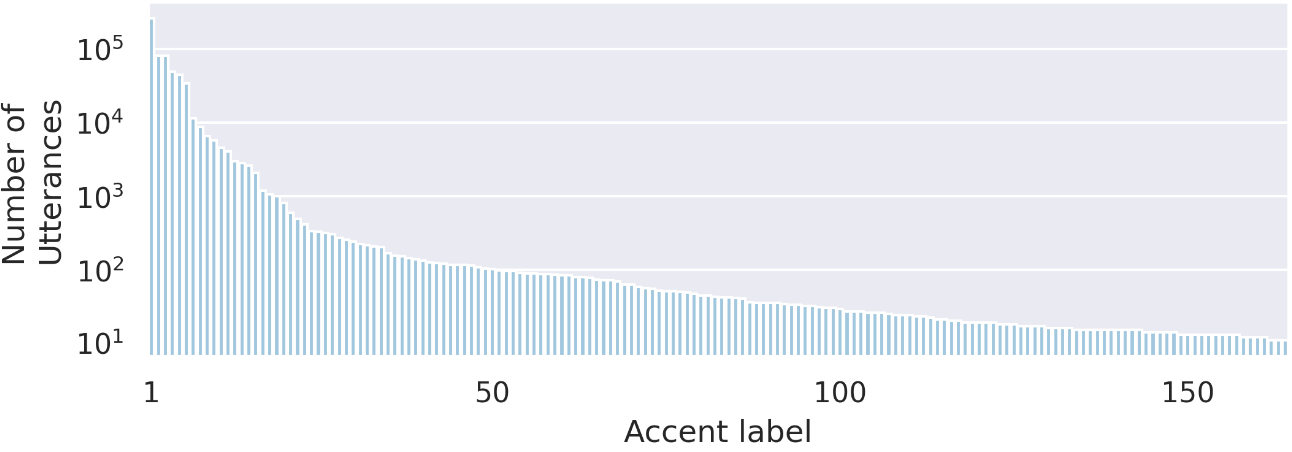}
\caption{Distribution of the number of utterances in different ground-truth accent labels \reg{in the GLOBE dataset.}}
\label{accent}
\end{figure}

\begin{figure}[t]
\centering
\includegraphics[width=0.4\textwidth]{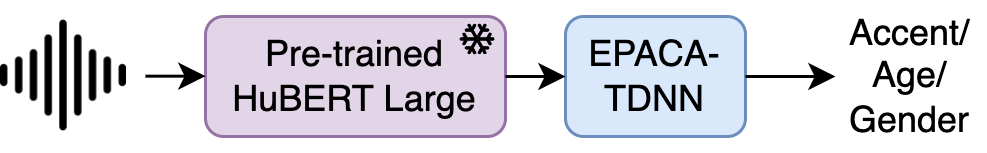}
\caption{Structure of speaker information prediction model, which is based on Hubert \cite{DBLP:journals/taslp/HsuBTLSM21} and EPACA-TDNN \cite{DBLP:conf/interspeech/DesplanquesTD20}.}
\label{hubert}
\vspace{-0.3cm}
\end{figure}

\blue{The final development stage involves populating missing metadata for speakers including accent, age and gender, and applying speech sample post-processing. Initially, accents represented by fewer than five speakers or 10 utterances are merged with the most similar accent label or removed from the dataset, as we believe that such limited speech samples do not adequately capture the full scope of an accent's characteristics. Furthermore, meaningless accent labels, such as ``not bad'' and ``A'lo,'' are also removed along with their corresponding speakers. Subsequently, for each metadata class, i.e., accent, age and gender, a training dataset, along with label-balanced validation and test sets, are developed from the refined speech data and speaker labels. Each training set includes utterances from 11,000 speakers, while each validation and test set features at least 1,000 speakers. Utilizing these subsets, three speaker information prediction models with the same structure, as depicted in Figure \ref{hubert}, are developed to predict the speaker’s accent, age and gender, respectively.} Due to the long-tail distribution of the speaker's accent label, \blue{as shown in Figure \ref{accent}}, the square-root sampling method \cite{DBLP:conf/iclr/KangXRYGFK20} is employed during model training to mitigate the negative impact. These models finally achieve accuracies of 97.22\%, 99.55\%, and 99.95\% for accent, age, and gender prediction, respectively, across the test sets. The models are then leveraged to populate the missing speaker metadata. After that, post-processing is applied to all utterances, which includes the elimination of leading and trailing silences based on VAD results and the \blue{further suppression of the background noise} through a speech enhancement tool \footnote{https://podcast.adobe.com/enhance}.

\section{Experiments}

\subsection{Experiments for Ground-Truth Speech Samples}

\subsubsection{Experimental setups}
\label{metrics}

In the first experiment, we evaluated the quality of ground-truth speech samples within GLOBE and other popular English multi-speaker datasets \cite{VCTK, DBLP:conf/icml/MinLYH21, DBLP:journals/corr/abs-2305-18802, DBLP:conf/lrec/ArdilaBDKMHMSTW20}. To objectively evaluate audio quality, we randomly selected 10,000 samples from the full set of VCTK \cite{VCTK}, the training set of GLOBE and Common Voice \cite{DBLP:conf/lrec/ArdilaBDKMHMSTW20} and the ``train-clean'' subsets of LibriTTS \cite{DBLP:conf/icml/MinLYH21} and LibriTTS-R \cite{DBLP:journals/corr/abs-2305-18802}. These subsets were selected because they represent the highest audio quality available in each dataset. For subjective evaluations, particularly the mean opinion score, we randomly chose 120 samples from those used in the objective evaluation for each dataset. The following evaluation metrics were utilized:

\textbf{Naturalness Mean Opinion Score (NMOS).} To evaluate the naturalness of speech samples, following \cite{DBLP:journals/corr/abs-2305-18802, DBLP:conf/nips/ArikCPPZ18}, we employed the Mean Opinion Score. Participants were asked to rate the naturalness of each utterance using a five-point Likert Scale \cite{DBLP:conf/icml/MinLYH21}. Each speech sample was rated by five distinct participants, and we calculated the average score along with a \(95\%\) confidence interval by the official tool \footnote{https://github.com/luferrer/ConfidenceIntervals} for each evaluated dataset.

\textbf{UTokyo-SaruLab Mean Opinion Score (UT-MOS) \cite{DBLP:conf/interspeech/SaekiXNKTS22}.} In line with prior studies \cite{DBLP:journals/taslp/HuangLLCL22, DBLP:conf/iscslp/XueDHLSL22}, predicted NMOS values were also provided for reference. The UT-MOS model was employed for NMOS prediction, which achieved state-of-the-art performance in 10 out of 16 metrics in the VoiceMOS Challenge \cite{huang22f_interspeech}.

\textbf{Word Error Rate (WER).} Following \cite{DBLP:journals/corr/abs-2305-18802, DBLP:conf/icml/MinLYH21}, we employed the WER metric to measure the average misalignment in speech transcripts relative to the ground-truth text. A lower WER indicates more accurate alignment. Speech transcription was conducted using a pre-trained Conformer-based automatic speech recognition model \red{\footnote{https://huggingface.co/nvidia/parakeet-rnnt-1.1b}} \cite{DBLP:journals/corr/abs-2305-05084}.

\textbf{Speaker Embedding Cosine Similarity (SMCS).} Consistent with \cite{DBLP:journals/corr/abs-2305-18802, DBLP:conf/icml/CasanovaWSJGP22}, speaker embedding cosine similarity was used to evaluate the similarity between two speeches from the same speaker. The speaker embeddings were extracted using the TitaNet-L speaker verification model \cite{DBLP:conf/icassp/KoluguriPG22}, which achieves the state-of-the-art equal error rate on the VoxCeleb1 \cite{DBLP:conf/interspeech/NagraniCZ17}.

\textbf{Speaker Embedding Vendi Score (SEVS).} Considering that the number of speakers in a dataset does not necessarily reflect its speaker or accent diversities, we introduced the speaker embedding-based Vendi Score \cite{friedman2023the} to evaluate the diversity of accents contained in each dataset. The Vendi Score is defined as the exponential of the Shannon entropy of the eigenvalues of a similarity matrix and has been used in both computer vision \cite{DBLP:journals/corr/abs-2312-03817} and natural language processing research \cite{DBLP:journals/corr/abs-2401-00690}.

\begin{table}[h]
\caption{Evaluation Results of GT Speech Samples}
\begin{adjustbox}{width=8cm,center}
\label{gtres}
\scriptsize
\begin{tabular}{c|ccccc}
\toprule
Corpus     & NMOS$\uparrow$   & UT-MOS$\uparrow$ & WER(\%)$\downarrow$ & \(SMCS\)$\uparrow$ &  SEVS$\uparrow$                   \\ \midrule
VCTK \cite{VCTK}      & \(4.23\pm0.06\)   &    4.01       &  \textbf{2.1}  &  0.887  &  71.16  \\ 
LibriTTS \cite{DBLP:conf/icml/MinLYH21} & \(4.20\pm0.07\)   &    4.00       &  3.8 &  0.893  &  94.71 \\ 
LibriTTS-R \cite{DBLP:journals/corr/abs-2305-18802}  & \(4.27\pm0.07\)   &    \textbf{4.12}       &  3.8 &  0.895  &  94.18 \\ 
Common Voice \cite{DBLP:conf/lrec/ArdilaBDKMHMSTW20}  & \(3.54\pm0.10\)   &    3.68       &  7.3   &  0.887 &  \textbf{112.21} \\ 
GLOBE       & \(4.25\pm0.06\)   &    4.09       &  3.9   &  \textbf{0.903}  &  110.08 \\ \bottomrule 
\end{tabular}
\end{adjustbox}
\end{table}

\begin{table*}[ht]
\setlength\tabcolsep{3pt}
\caption{Evaluation Results of the modified YourTTS Trained on Different Datasets.}
\begin{adjustbox}{width=16cm,center}
\begin{threeparttable}
\label{basemodel}
\scriptsize
\begin{tabular}{c|ccccc|ccccc}

\toprule
 Evaluation Corpus               & \multicolumn{5}{c|}{\(\rm LibriTTS_{test}\)}         &\multicolumn{5}{c}{\(\rm GLOBE_{test}\)}\\ \midrule
 Metrics  & NMOS$\uparrow$  & SMOS$\uparrow$    & SMCS$\uparrow$                      & WER(\%)$\downarrow$ & UT-MOS$\uparrow$   & NMOS$\uparrow$    & SMOS$\uparrow$     & SMCS$\uparrow$                    & WER(\%)$\downarrow$    & UT-MOS$\uparrow$             \\ \midrule
 Ground Truth  & \(4.16\pm0.06\) & \(4.02\pm0.05\)    & 0.772  & 3.82 & 3.93  & \(4.19\pm0.06\)  & \(4.08\pm0.08\) & 0.774   & 3.91  & 4.07      \\ \midrule
 VCTK \cite{VCTK}  & \(3.91\pm0.07\) & \(3.72\pm0.08\)    & 0.716  & 6.4 & 3.76  & \(3.89\pm0.08\)  & \(3.52\pm0.09\) & 0.698   & 6.6   & 3.78      \\
 LibriTTS \cite{DBLP:conf/icml/MinLYH21} & \(4.03\pm0.07\) & \(3.83\pm0.07\) & \textbf{0.740}  & \boldmath\(6.2\) & 3.88  & \(4.01\pm0.07\)  & \(3.63\pm0.09\)  & 0.715  & 6.4  & 3.83      \\
 LibriTTS-R \cite{DBLP:journals/corr/abs-2305-18802} & \(4.09\pm0.07\) & \(3.80\pm0.08\)  & 0.736  & 6.3 & \textbf{3.91} & \(4.02\pm0.06\)  & \(3.62\pm0.09\)  & 0.712  & 6.4 & 3.85  \\
 Common Voice \cite{DBLP:conf/lrec/ArdilaBDKMHMSTW20} & \(2.98\pm0.07\) & \(3.79\pm0.09\)  & 0.733  & 10.6 & 2.86 & \(3.07\pm0.07\)  & \(3.72\pm0.08\)  & 0.726  & 10.3 & 2.97  \\
 GLOBE         & \(4.03\pm0.06\) & \(3.84\pm0.07\)  & 0.738  & 6.3 & 3.85 & \(4.05\pm0.08\) & \(3.81\pm0.08\) & \textbf{0.732}   & \textbf{6.3} & \textbf{3.86}  \\ \bottomrule
\end{tabular}
\end{threeparttable}
\end{adjustbox}
\vspace{-0.3cm}
\end{table*}

\begin{figure}[h]
\centering
\includegraphics[width=0.35\textwidth]{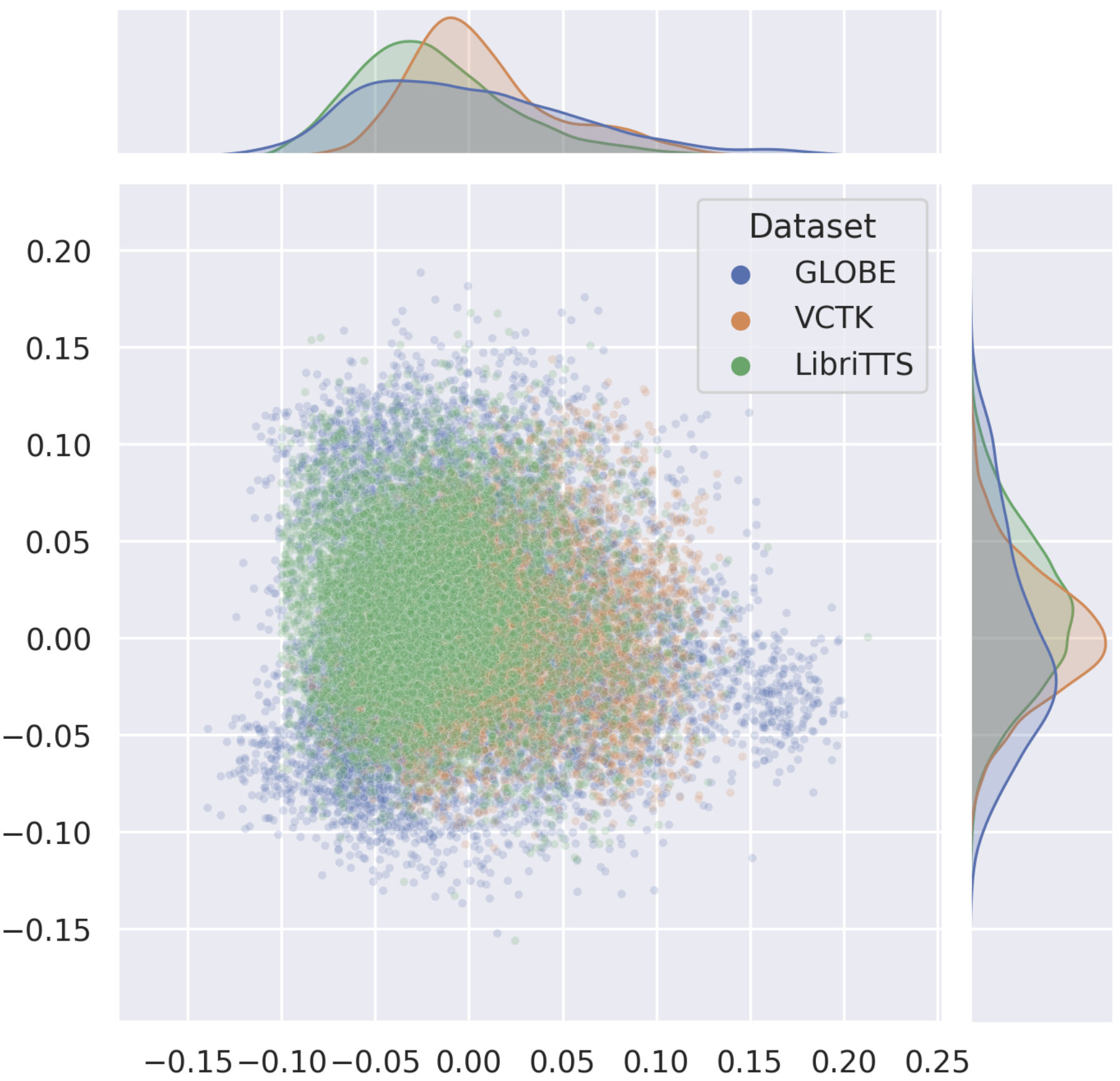}
\caption{Speaker embedding distributions and their marginal distributions KDE of different datasets after dimension reduction via PCA.}
\label{dist}
\end{figure}

\subsubsection{Experimental results}

Table \ref{gtres} presents the experiment results for ground-truth speech samples. Concerning speech naturalness, all datasets, except for the Common Voice dataset, which recorded a lower NMOS of 3.54 due to prevalent low-quality audio, achieved similar NMOS scores. We also conducted the Mann–Whitney–Wilcoxon (MWW) test \cite{wilcoxon1992individual} to determine whether there are statistically significant differences in NMOS scores between any two datasets. It was found that the \(p\)-value between LibriTTS and GLOBE was \(0.91 > 0.05\), indicating that the speech samples from GLOBE do not show a statistically significant difference from LibriTTS in naturalness. In contrast, the \(p\)-value between Common Voice and GLOBE is \(4.3e^{-6} < 0.05\), denoting that the naturalness of speech from GLOBE is statistically better than that of Common Voice. These findings are also corroborated by UT-MOS results. Regarding the WER, the VCTK dataset had the lowest WER, with LibriTTS, LibriTTS-R, and GLOBE showing comparably low WER levels, contrasting with Common Voice's higher WER. For SMCS, all datasets displayed similar scores, suggesting well-defined speaker characteristics within each dataset, given that TitaNet-L's threshold for determining utterances from the same speaker is 0.7. SEVS results indicate that LibriTTS and LibriTTS-R have wider speaker diversity than VCTK, with both Common Voice and GLOBE also showing improvements in speaker diversity compared to LibriTTS and LibriTTS-R. In summary, these results demonstrate that GLOBE achieves speech naturalness comparable to other popular TTS datasets by filtering out low-quality speech samples from Common Voice while maintaining richer speaker diversity compared to other datasets. \blue{We also visualized the speaker embedding distributions and their marginal distributions' kernel density estimates (KDE) of the GLOBE, VCTK, and LibriTTS datasets, as shown in Figure \ref{dist}. To do this, we extracted 10,000 speaker embeddings from the samples that were utilized for objective evaluation and reduced the dimension of all speaker embeddings to 2 via principal component analysis (PCA). As illustrated in the figure, the GLOBE dataset, represented in blue, displays a broader distribution of speaker embeddings across both dimensions compared to the other datasets, indicating a wider variety of speaker characteristics contained within the GLOBE corpus.}

\subsection{Experiments for Speaker Adaptive TTS Synthesized Samples}

\subsubsection{Model details}
\blue{To investigate the influence of training datasets on the synthesized speech quality of zero-shot speaker-adaptive TTS models, we employed YourTTS \cite{DBLP:conf/icml/CasanovaWSJGP22}, \rec{a widely used zero-shot speaker-adaptive TTS approach}, as the baseline model with three modifications: firstly, the language embedding was removed from the model, given our focus solely on English; secondly, to thoroughly assess the datasets' influence on model performance and avoid bias introduced by pre-trained models, we replaced the pre-trained speaker encoder with a trainable encoder, i.e., EPACA-TDNN \cite{DBLP:conf/interspeech/DesplanquesTD20}; thirdly, some model parameters were adjusted to facilitate training with 24 kHz audio data.}




\subsubsection{Experimental setups}
We trained the modified YourTTS models on all datasets outlined in Section \ref{datasets}. For the VCTK dataset, training encompassed the entire dataset. For both the LibriTTS and LibriTTS-R datasets, training was performed on the ``train-clean'' and ``train-other'' subsets. In terms of the Common Voice and GLOBE datasets, models were trained on the training subset. Throughout the training phase, we downsampled all speech samples to a 24 kHz sampling rate. Both phoneme sequences used for training and evaluation were generated from the ground-truth text using Phonemizer \cite{Bernard2021}. Training for both models was executed end-to-end and each training session was conducted on two NVIDIA V100 GPUs. The YourTTS model was trained for \(1.8m\) iterations with a total batch size of 48. All training utilized AdamW optimizer \cite{DBLP:conf/iclr/LoshchilovH19}, featuring $\beta_1=0.8$, $\beta_2=0.99$, and a weight decay parameter of 0.01. The initial learning rate was set at $2\times10^{-4}$ and followed a decay factor of $\gamma=0.9999$. 

For evaluation, we employed several metrics introduced in Section \ref{metrics}. Specifically, we assessed the naturalness of the synthesized speech for each model using NMOS and UTMOS. The intelligibility of the synthesized speech was evaluated using WER. Additionally, SMCS was used to measure the similarity between the synthesized speech and the ground-truth speech. Furthermore, we introduced an additional evaluation metric:
\textbf{Speaker Similarity Mean Opinion Score (SMOS) \cite{DBLP:conf/nips/ArikCPPZ18}.} Parallel to the NMOS, this metric evaluates the speaker similarity between the synthesized utterance and a random utterance from the same speaker. Assessments were conducted on a five-point Likert Scale \cite{DBLP:conf/icml/MinLYH21} by the same participants of the NMOS. The mean score and confidence interval were also calculated.

\subsubsection{Experimental results}

Table 3 presents the evaluation results for the modified YourTTS model trained on different datasets. When evaluated on the LibriTTS test set, the model trained on LibriTTS-R achieved the highest NMOS, closely followed by models trained on GLOBE and LibriTTS datasets, which exhibited no statistically significant difference in NMOS, as indicated by a \(p\)-value \(> 0.05\) in the MWW test. The model trained on the Common Voice dataset underperformed, exhibiting a statistically significant decrease in NMOS, with the \(p\)-value in the MWW test at \(3.2e^{-4}\), which is \( < 0.05\). In terms of SMOS, models trained on LibriTTS, LibriTTS-R, Common Voice, and GLOBE yielded comparable outcomes. However, the model trained on VCTK performed slightly worse. This discrepancy can be attributed to the dataset's limited speaker diversity, which also adversely affected the SMCS scores. In the evaluation of the GLOBE test set, NMOS results mirrored those obtained on the LibriTTS test set. However, a significant decline in SMOS was observed across all models, attributed to the GLOBE set's broader diversity of speaker accents, presenting a notable challenge to the models' generalization capabilities. Specifically, the SMOS for the baseline model trained on LibriTTS dropped by 0.20 compared to its performance on the LibriTTS test set, with statistically significant \(p\)-values at \(9e^{-4} < 0.05\) from the MWW test. In contrast, models trained on GLOBE exhibited the smallest decline in speaker similarity, and the MWW test indicated that these declines were not statistically significant, with \(p\)-values of \(0.54 > 0.05\), demonstrating that the broad accent coverage by GLOBE enhanced TTS models' adaptability to diverse accents. In summary, these results demonstrate that the speaker-adaptive TTS model trained on GLOBE exhibits better generalization compared to models trained on other TTS datasets, enabling it to effectively adapt to various accents.

\section{Conclusions}

This paper introduces GLOBE, a high-quality English corpus featuring worldwide accents originating from Common Voice, aimed at addressing the poor generalizability issue of current zero-shot speaker-adaptive TTS models. GLOBE not only matches the audio quality of popular TTS datasets like LibriTTS but also surpasses them by covering a broader range of worldwide accents and offering metadata for an extensive array of over 20,000 speakers. Our experiments demonstrate that speaker-adaptive TTS models trained on GLOBE achieve better generalizability than those trained on other datasets. We hope that the release of GLOBE will contribute to advancements in TTS research.

\newpage
\pagebreak

\bibliographystyle{IEEEtran}
\bibliography{mybib}

\end{document}